\documentclass[pre,twocolumn,amsmath,amssymb,floatfix,superscriptaddress]{revtex4-2}

\usepackage{graphicx}
\usepackage{dcolumn}
\usepackage{bm}
\usepackage{booktabs}

\usepackage{lipsum}
\usepackage{cancel}
\usepackage[normalem]{ ulem }
\usepackage{soul}
\usepackage[FIGTOPCAP,tight,raggedright,nooneline]{subfigure}
\usepackage{float}
\graphicspath{{images/}{plots/}}
\usepackage{hyperref}

\newcommand{\avg}[1]{\ensuremath{\left< #1 \right>}}

\usepackage{xcolor}

\definecolor{applegreen}{rgb}{0.55, 0.71, 0.0}

\begin{document}

    \title{Higher order interactions destroy phase transitions in Deffuant opinion dynamics model}

    \author{Hendrik Schawe}
    \email{hendrik.schawe@cyu.fr}
    \affiliation{Laboratoire de Physique Th\'{e}orique et Mod\'{e}lisation, UMR-8089 CNRS, CY Cergy Paris Universit\'{e}, France}
    \author{Laura Hern\'{a}ndez}
    \email{laura.hernandez@cyu.fr}
    \affiliation{Laboratoire de Physique Th\'{e}orique et Mod\'{e}lisation, UMR-8089 CNRS, CY Cergy Paris Universit\'{e}, France}

    \date{\today}

    \begin{abstract}
        
        We define a higher order Deffuant model  by generalizing the original  pairwise interaction 
         model for bounded-confidence opinion-dynamics to interactions involving a   group of agents of size $k$. The generalized model is naturally encoded in a hypergraph. We  study this dynamics in different hypergraph topologies, from random  hypergraph ensembles, to spatially embedded hyper-lattices. We show that including higher order interactions induces a  drastic change in the onset of consensus for random hypergraphs; instead of the sharp phase transition, characteristic of the dyadic Deffuant model, the system undergoes a smooth size independent crossover to consensus, as the confidence value increases. This phenomenon is absent from regular hypergraphs, which conserve a phase transition.  
    \end{abstract}

    \maketitle

    \section{Introduction}

        The formation and diffusion  of opinion in societies have largely been studied from the point of view of Physics, modeling the society in terms of  a dynamical system of interacting agents, by the means of stylized agent based models, aimed at understanding the role played by different aspects of social interaction in the observed patterns in real life. Different models have been proposed which may be classified according to the representation of the agents' opinions (scalar or vector, continuous or discrete variables), or by the structure of their interactions (mixed population, or networked systems), or even by the detailed aspects of the dynamics which are in generally grounded on disciplinary knowledge issued from social sciences, like for example, the Social Influence Theory~\cite{Castellano2009Statistical,kelman1958compliance}.
        
        The most popular aspects of social interactions, widely considered by previous studies are
        \emph{homophily}, agents interact preferably with similar agents, and \emph{social influence}, agents which interact become more similar. One influential class of opinion dynamics models are \emph{bounded confidence} models, which implement homophily by a threshold rule: only agents  whose opinions lie within a \emph{confidence} range may interact. Two outstanding models of this class are the Deffuant-Weisbuch (DW) model \cite{deffuant2000mixing} and the Hegselmann-Krause (HK) model \cite{hegselmann2002opinion}. Both model the opinion of the $N$ agents in the population as a continuous variable $x_i\in [0,1]$, $\forall i=1,N$ and their main difference is that while the DW considers pairwise interactions and asynchronous updates, in the HK model, at each step, all the agents synchronously update their opinion by taking the average of each agent's current opinion and those of their neighbors.

        All these models consider pairwise relations between agents which can naturally be modeled by networks \cite{Castellano2009Statistical}. However,  pairwise interactions do not describe all possible ways of discussion in real life, and the particularities of group discussion and decision making are still a matter of discussion in Social Psychology~\cite{levendusky2016group,kerr2004group,binder2006direct}. This necessity of going beyond pairwise interactions has been first addressed by generalizing previous models to the case of \textit{group interactions} mainly in the form of a majority rule, as in the \textit{voter model}~\cite{redner2019reality} or in the form of an aggregation rule that averages the opinion of the neighbours of the active agent, as in the Hegselmann-Krause model~\cite{hegselmann2002opinion}.
        
        Recently, the interest in multi-agent interactions \cite{lambiotte2019networks} to model group dynamics at a large scale did rise sharply and several studies were published in the context of opinion dynamics \cite{maletic2014consensus,horstmeyer2020adaptive,neuhauser2020multibody,cencetti2020temporal,sahasrabuddhe2021modelling,noonan2021dynamics,hickok2021bounded}, (social) contagion \cite{iacopini2019simplicial,landry2020effect,barrat2021vanishing} and other dynamical processes \cite{arruda2021phase,guo2021evolutionary}, which modeled the topology of interactions by hypergraphs.

        It has been shown that if the multi-agent interaction is non-linear, -\textit{higher order interaction} (\textit{HOI})-, the system cannot be modeled by any inherently pairwise graph \cite{neuhauser2020multibody}. Regarding opinion dynamics, multi-agent interactions need to be considered to address problems where individuals discuss in groups, like professional meetings or private instant messenger groups.

        Here, we generalize the DW model to the case  where the interactions occur in small groups. We perform extensive simulations of this \textit{higher order interaction Deffuant model}, (HOID), for hypergraphs of different topology. Our results  show that considering hyperedges of size $k=3$ is already enough to modify qualitatively         the way consensus is reached with respect to the outcomes of the pairwise DW dynamics. This modification is enhanced with the introduction of larger hyperedges to the extent that the phase transition from polarization to consensus is replaced by a smooth crossover. We also show that, as for networks, the outcomes of the dynamics strongly depend on whether the hypergraph is regular or random.

        While preparing this manuscript, we noticed Ref.~\cite{hickok2021bounded} appearing on a preprint server, which also studies a generalization of the DW model to HOI although with a different model and approach. 
        In Ref.~\cite{hickok2021bounded}, the strict \emph{bounded confidence} rule of the dyadic DW model is replaced by a smoother criterion which enhances the interaction of large hyperedges.         Here, on the contrary, we  simply extend the standard notion of bounded confidence to the hyperedge: the group will only interact if all members of a hyperedge are within the confidence interval of each other, which naturally brings as a consequence, that large hyperedges are less likely to interact than smaller ones. Moreover Ref.~\cite{hickok2021bounded} mainly studies the mixed population case, along with some simulations of sparse hypergraphs of very small size, while we focus in exploring the interplay of the dynamical rules and the interaction structure, with a particular attention in the finite size effects that have been shown to be dominant in bounded confidence models in networks~\cite{schawe2021bridges}.  
        Interestingly,  these two  complementary works, address two possible alternative situations: while  Ref.~\cite{hickok2021bounded}, assumes the existence of  nodes that could reduce discordance in the group, here we focus on  the role of a blocking minority.

    \section{Models and Methods}
    \label{sec:model_methods}
        We propose the \emph{higher order interaction Deffuant} model (HOID), which generalizes the Deffuant-Weisbuch (DW) model \cite{deffuant2000mixing} from pairwise  to higher order interactions. This generalization of the DW model is straightforward because, unlike the HK model where node variables are synchronously updated, in the DW  model, the updates are done at the level of a binary edge, which is easily generalized to the update of hyperedges.
        
        The original DW model is defined for a set of $N$ agents each with a continuous opinion $x_i \in [0, 1]$. The agents can interact pairwise, provided that the difference of their opinions lies  within a \emph{confidence interval} given by an external parameter $\varepsilon$, and they can also be restricted by an underlying network (e.g., a  lattice or a random graph). The asynchronous dynamics takes place in discrete time and at every time step, a pair of neighboring agents $i$ and $j$ attempts to interact and update their opinion according to

  \begin{align}
         x_i(t+1) = \begin{cases}
                (x_i(t) + x_j(t)) / 2, & \text{if } |x_i - x_j| < \varepsilon\\
               x_i(t), & \text{otherwise}
            \end{cases}
         \end{align}       
        
        Note that this dynamics is a particular case of the original model, where the amplitude of opinion change towards the mean opinion is given by a parameter, $\mu$. In our case, this parameter is set to its maximal value, such that both agents assume their average opinion after one successful interaction. For homogeneous confidences this should result in a higher convergence speed to the final state. 
        This update rule means that either two neighbors discuss and  arrive at a compromise opinion or do not discuss at all, depending on the confidence parameter $\varepsilon$.

        To account for the fact that discussions are not exclusively happening between two persons, but may involve a group of agents, we need to replace graphs encoding pairwise relations, by hypergraphs.
                A \emph{hypergraph} $\mathcal{H} = (V, E)$ is defined by a set of nodes or vertices $V$ representing the agents and a set of hyperedges $E$, which is a subset of the powerset of $V$, i.e., can contain any subset of $V$.
        This way a hyperedge $e \in E$ establishes a relation between its members, which encodes the group interaction. The number of nodes $N = |V|$ is called the \emph{size} of the hypergraph or the \emph{system size}. The \emph{degree} $d_i$ of a node $i$ is the number of hyperedges the node is a \emph{member} of.
        A hypergraph is called \emph{uniform} or \emph{$k$-uniform}, if all hyperedges $e \in E$ have the same \emph{size} $k = |e|$.
        For clarity, we refer to conventional graphs as \emph{dyadic graphs} or \emph{$2$-uniform hypergraphs}.

        In this framework we modify the dynamical rule such that at each time step a random hyperedge $e$ is selected
        and every member $i \in e$ is updated according to
        \begin{align}
            x_i(t+1) = \begin{cases}
                \overline{x_e}, & \text{if } \max_{j \in e}{x_j(t)} - \min_{j \in e}{x_j(t)} < \varepsilon\\
                x_i(t), & \text{otherwise},
            \end{cases}
        \end{align}
        where $\overline{x_e} = \frac{1}{|e|} \sum_{j \in e} x_j$ is the average opinion of all members of the hyperedge. So interaction only happens if all members are within the confidence range of each other. 
        This rule addresses the situation where an individual holding a very different opinion from the rest of the group, by blocking the discussion, prevents an otherwise possible compromise  to reach consensus.

        For small confidence ranges $\varepsilon$ this means that the probability of a successful interaction for a hyperedge (provided that the opinions of the members are random and independent, which is the case for the initial conditions of our model) decays exponentially in the size $k$ of the hyperedge, and therefore, large groups have a low probability to reach a compromise opinion.

\begin{figure}[htb]
            \centering
            \includegraphics[scale=0.7]{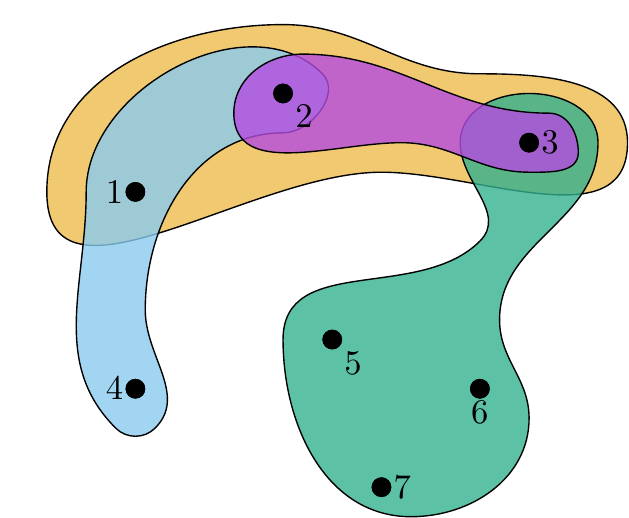}
            \includegraphics[scale=0.7]{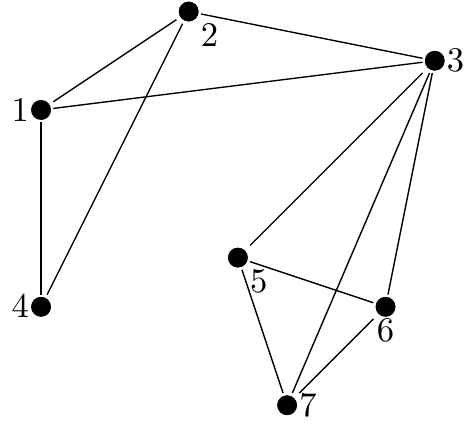}
            \caption{\label{fig:graph_ex}
                Example of a hypergraph (left) and its projection to a dyadic graph. In the hypergraph, the degree of node $4$ is $d_4 = 1$, since it is only member of one hyperedge, the degree of node $2$ is $d_2 = 3$ since it is a member of three hyperedges.
            }
        \end{figure}

        It is worthwhile noticing that the group interaction proposed here is different from the interaction between an agent and its group of neighbors that rules the dynamics of the Hegselmann-Krause (HK) model \cite{hegselmann2002opinion}. This becomes apparent when considering the projection of an example hypergraph onto a dyadic graph shown in Fig.~\ref{fig:graph_ex}. In the HK model, each agent checks every neighbor synchronously and updates its own opinion by taking into account the opinion of all its neighbours whose opinion differs from its own in less than the confidence, $\varepsilon$, 
         regardless of the differences between the opinions of those neighbours among themselves (which could be larger than the confidence). The HOID model, on the other hand, updates one hyperedge at a time. This means that in the example of Fig.~\ref{fig:graph_ex}, the state of vertex 3 might be updated three times (when updating hyperedges green, violet and orange) provided that all the nodes of each hyperedge have their opinions within the confidence range. As a consequence, a dissenter can block the interaction of all other agents in the hyperedge, a mechanism absent from the HK model.
        Indeed, if the update rule is non-linear, like the threshold value for the HOID, hypergraph interactions cannot be mapped to a dyadic graph.
        Also note that for a $2$-uniform hypergraph this model reduces directly to the well studied DW case.

        As usual, we start the dynamics with initial opinions drawn from a uniform distribution $U[0, 1]$. The dynamics eventually leads the system, after a long transient, to a \emph{final state} \cite{lorenz2005stabilization}, where the opinions of the agents do not change anymore. 
               As convergence criterion we require that after performing a \emph{sweep}, i.e., $N$ attempted updates, we have:
        \begin{align}
        \label{eq:criterion}
            \sum_{i=0}^{N} |x_i(t) - x_i(t-1)| < 10^{-3}
        \end{align}
        This criterion, which has already been tested in different studies of the HK model \cite{schawe2020open,schawe2020cost,schawe2021bridges} does become sharper for larger systems and is especially suited to ensure that regions of very dense agents are static. Since our main observable is the \emph{relative size of the largest cluster} $S$, i.e., the fraction of agents having the majoritarian opinion within a tolerance of $10^{-3}$, we expect this criterion to lead to accurate results while saving a lot of computations on converged systems.

    \subsection{Random hypergraph ensembles}

        Although the original DW model was also studied  on  a complete graph, its generalization to   HOI in this configuration is less interesting. In fact,  the  interaction rule         applied here induces a very low probability for the         large hyperedges  to interact. As the complete hypergraph contains far more large hyperedges than small ones, this will lead to blocking the evolution of the system,  making the complete hypergraph non interesting.  Instead, we focus on a selection of ensembles of \emph{sparse} hypergraphs, i.e., ensembles whose mean degree is independent of the number of nodes. For the sake of comparison, we will generalize the HOID model to lattices and random graph ensembles, on which the DW \cite{stauffer2004simulation,weisbuch2004bounded,fortunato2004universality} and related models \cite{schawe2021bridges} were studied before. 

    \subsubsection{Erd\H{o}s-Reny\'i}
        First, we consider a generalization of the Erd\H{o}s-Reny\'i (ER) ensemble to hypergraphs. The dyadic ER consists of graphs where every edge exists with probability $p$. In the limit of large graphs, $p = c / N$, where $c$ is the finite expected degree the graphs are sparse.
        
        To construct hyperedges in the same way, let us call $p_k$, the probability that $k$ nodes taken at random, constitute  an hyperedge.
        Therefore 
        \begin{align}
            \label{eq:er_c}
            c_k = p_k \frac{k}{N} \binom{N}{k}
        \end{align}
        is the expected degree contributed by hyperedges of size $k$. The total expected degree is simply $c = \sum_k c_k$.

        Technically, we construct realizations of this ensemble by first determining how many $k$-hyperedges should appear in the graph by drawing a binomially distributed random number from $\mathcal{B}(\binom{N}{k}, p_k)$, since it is infeasible to iterate all $\binom{N}{k} = \mathcal{O}(N^k)$ possible hyperedges of size $k$ and decide whether to include them or not with probability $p_k$. Since we are interested in sparse hypergraphs, we calculate $p_k$ from the desired mean degree using the relation given by Eq.~\eqref{eq:er_c}. For large values of $N$ it becomes impractical to draw the number of hyperedges from a binomial distribution. Therefore, for an expected number of hyperedges $M_k = \binom{N}{k} p_k = N c_k / k > 10^3$,  we switch to the Gaussian $\mathcal{N}(M_k, \sqrt{M_k(1-p_k)p_k})$, ensuring that the error introduced by this approximation is negligible.

    \subsubsection{Barab\'{a}si-Albert}
        Additionally, we introduce a scale-free $k$-uniform hypergraph, in the sense that the degree distribution   -the number of hyper-edges a node belongs to-  follows a power-law with an exponent $2 < \gamma \le 3$. To construct a realization we perform the preferential attachment procedure of the Barab\'{a}si-Albert (BA) graph ensemble \cite{barabasi1999emergence} with hyperedges, which is a special case of the ensembles introduced in \cite{avin2019preferential,giroire2021preferential}. This ensemble offers the parameter $m$ determining the number of hyperedges introduced for each node, which therefore determines the average degree %$\avg{d} = c \approx m k$.
        $c \approx m k$.
        We start with a fully connected \emph{core}, i.e., all subsets of size $k$ are hyperedges, of $M = \max(m-1, k)$ nodes. The remaining nodes are iteratively added. For each node $m$ hyperedges are introduced and their other $k-1$ neighbors are chosen as members with a probability proportional to their current degree, avoiding identical hyperedges and nodes appearing twice in the same hyperedge.
        This procedure leads to a scale free degree distribution $P(d) \propto d^{-\gamma}$ with $\gamma = 2+\frac{1}{k-1}$ \cite{avin2019preferential} and reduces to the well known BA case for $k=2$.

    \subsection{Regular spatial hypergraphs}
    
        The DW model was also studied is the square lattice as a very stylized way to introduce a neighbourhood embedded in real space  \cite{deffuant2000mixing,weisbuch2002meet}. 
        
\begin{figure}[htb]
            \centering
            \includegraphics[scale=0.7]{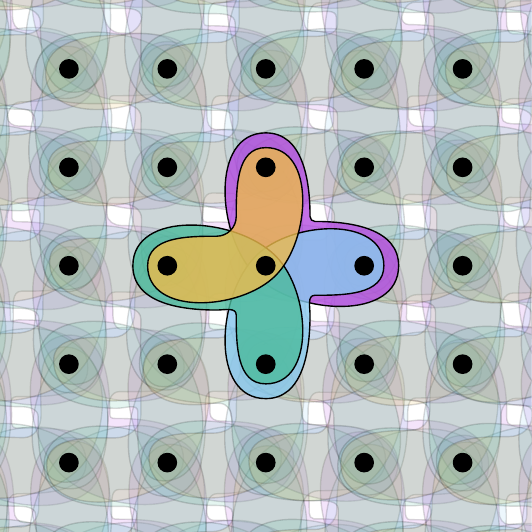}
            \includegraphics[scale=0.7]{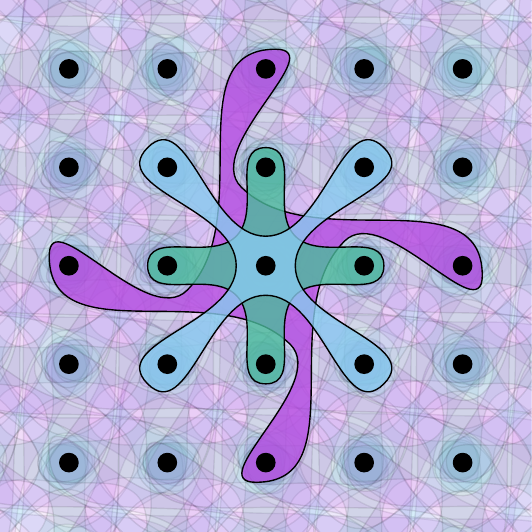}
            \caption{\label{fig:lattice_ex}
                Two possible hypergraph configurations for lattices with
                (a) $k=3, c=12$
                (b) $k=5, c=15$.
                For clarity, a `basis set' of hyperedges is shown for the central node in the foreground. Each node contributes such a basis set to the hypergraph. In the background, with muted colors, all hyperedges are drawn.
            }
        \end{figure}

        There are different ways to introduce HOI in lattices. Here
        we propose two regular structures based on the successive neighbours (first, second, third, etc. nearest neighbours)  of the dyadic square lattice.
        The first case is a $3$-uniform hypergraph, where hyperedges connect up to second nearest neighbors as shown on the left of Fig.~\ref{fig:lattice_ex}. This results in a mean degree of $c=12$.
        The second case is a $5$-uniform hypergraph, where hyperedges connect up to third nearest neighbour nodes as shown on the right of Fig.~\ref{fig:lattice_ex}. This results in a mean degree of $c=15$.

    \section{Results}
    \label{sec:results}
        Unless stated otherwise we perform simulations for $1000$ independent realizations of the system for each of $300$ equidistant values of $\varepsilon \in [0, 0.6]$, i.e., a resolution in $\varepsilon$ space of $0.002$. For better visibility we present the results as lines instead of symbols. The statistical uncertainty is generally of the order of the width of the line. The raw data of the final states consisting of the locations and sizes of all clusters and convergence times are openly available at \cite{rawDataHBC} for the $k>2$ cases.
        
        Note that the HOID on $k$-uniform hypergraphs reduces to the DW model on the corresponding topology for $k=2$, therefore a look back at the DW helps to identify the patterns that are directly related to the higher order interactions. We  include, for comparison, the results of the dyadic DW model  for each hypergraph topology.
        
       As explained before, we do not intend to explore the HOID model in the mixed population, due to the high proportion of blocked hyperedges, however a careful study of the dyadic DW model in the complete graph is useful to identify which observed phenomena are already present in the DW model in the complete graph, which are  induced by the interplay between the dyadic DW model and the underlying topology,  and finally, which are associated with the introduction of HOID.  Such study is included in  the Supplementary Material  where we revisit the results of the DW model for the complete network presented in \cite{schawe2021bridges}, including an extensive finite size study that goes well beyond the sizes considered so far. The corresponding raw data is openly available at \cite{rawDataDW}. 
 
    \subsection{Random Hypergraph Ensembles}
    \label{subsec:RHE}
    \subsubsection{Uniform Hypergraphs}
    \label{subsubsec:uniform}
    \paragraph{Erd\H{o}s-R\'{e}nyi Hypergraphs}

        \begin{figure}[htb]
            \centering
            \includegraphics[scale=1]{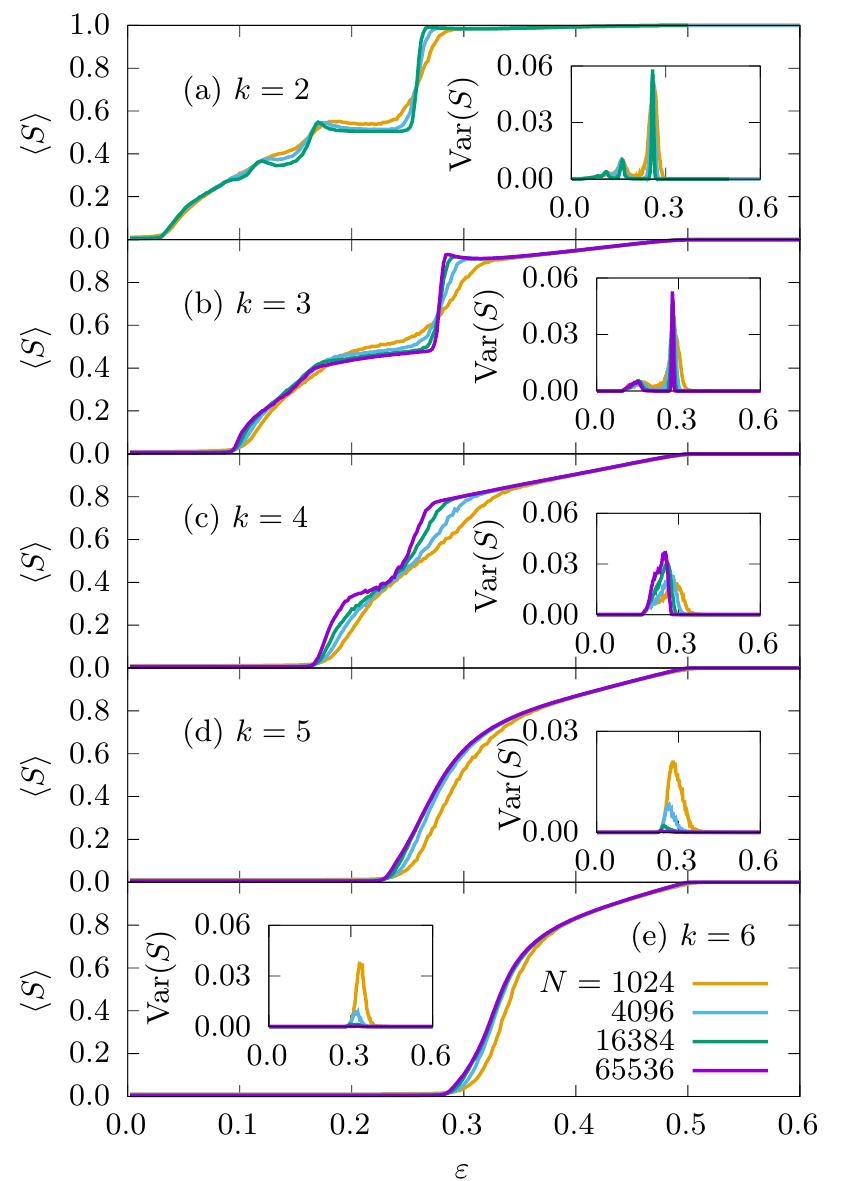}
            \caption{\label{fig:hbc_er10_s}
                Mean relative size of the largest cluster $\avg{S}$ as a function of the confidence $\varepsilon$ for the HOID model on $k$-uniform hyper-ER graphs with edge sizes $k\in \{2, 3, 4, 5, 6\}$ and an expected mean degree of $c=10$ for different system sizes each. The insets show the variance $\mathrm{Var}(S)$, which is sharpening for $k \le 3$ and vanishing for $k \ge 5$, supporting the change from a sharp transition to a crossover.
            }
        \end{figure}
        In this section we compare the behavior of the HOID model on $k$-uniform ER-hypergraphs, for different values of $k$. We are interested in relatively sparse hypergraphs so, unless stated otherwise, we consider $c=10$.

        The first striking effect of the HOID model is the region of total fragmentation with $\avg{S} \approx 0$  for low $\varepsilon$, which grows with $k$ in Fig.~\ref{fig:hbc_er10_s}. This phenomenon can be understood by recalling that, for fixed low values of $\varepsilon$, the probability for the agents connected by a hyperedge to interact shrinks exponentially in the number of members of the hyperedge, $k$, since the $k$ random initial opinions must lie within a range of $\varepsilon$. So the amount of blocked hyperedges grows with $k$ and inhibits the dynamics of the system for low values of $\varepsilon$.

        More interestingly the very existence of a phase transition, for sparse hypergraphs, also depends on $k$. Figure~\ref{fig:hbc_er10_s} shows that for $k=2, 3$ ($k=4$ behaving as a limit case) there is a transition from polarization to consensus which gets sharper with increasing system size, however for $k>4$ the transition disappears letting place to a crossover behaviour that becomes independent from the system size. The variances at the inset of the panels confirm this: while they are sharpening with system size for $k \le 3$, (in the manner of a diverging susceptibility) they vanish with system size, for $k \ge 5$

\begin{figure}[htb]
            \centering
            \includegraphics[scale=1]{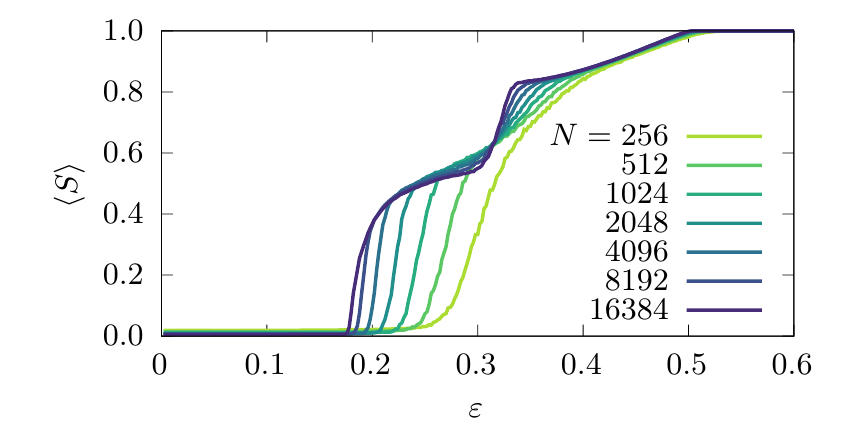}
            \caption{\label{fig:hbc_er150_s}
                Mean relative size of the largest cluster $\avg{S}$ as a function of the confidence $\varepsilon$ for the HOID model on $6$-uniform hyper-ER graphs with an expected mean degree of $c=150$ for different system sizes.
            }
        \end{figure}

        However, this behaviour is a characteristic of sparse hypergraphs. For very large  values of the average connectivity $c$, a sharp transition reappears  for the $6$-uniform ER-hypergraphs as shown in Fig.~\ref{fig:hbc_er150_s}.

        Here we can observe  once again, the importance of the size effects in the study of bounded confidence models, which require extensive simulations in order to reveal important qualitative aspects of their behaviour, as seen in~\cite{schawe2020open,schawe2021bridges}. Note that the curves for $N=256,512$ look very similar to the bottom panels of Fig.~\ref{fig:hbc_er10_s} even at the very high connectivity of $c=150$, it is necessary to go to larger sizes to observe the qualitatively different behavior. 
       
        Finally, as consensus is setting in with increasing $\varepsilon$, an extremely shallow minimum appears in the $\avg{S} $  curves, in the cases  $k=2,3$, for $\varepsilon \gtrsim \varepsilon_c$. This effect  is neither related to the HOI, not to the networked structure,  but is a consequence of  the asynchronous DW dynamics and is present in the DW model in the mixed population (see Supplementary Material). For larger $k$ this shallow valley is replaced by an almost perfectly linear increase in $\avg{S}$, before the onset of unanimity, for $\varepsilon=0.5$. This behaviour results from the interplay of the DW dynamics and the  hypergraph, and is absent from DW on networks. A heuristic argument allows to explain this behaviour: at these relatively high values of the confidence the majority of agents have converged to the consensus opinion and those who have not, are blocked by at least one \emph{blocking} agent in all the hyperedges to which they belong (notice that the other agents in the hyperedges could have already converged). Since each agent is part, on average, of $c=10$ hyperedges, it is probable that the agents who have not converged are themselves the blocking ones. This means that their opinion differs in at least $\varepsilon$ from the consensus opinion, where most of their neighbours are. Assuming that those blocking agents did not allow their edges to interact (or just few times) they are still  very close to their uniformly distributed initial opinion, and therefore their number will decrease linearly in $\varepsilon$,  which induces the linear growth in $\avg{S}$. 
        
        We will see later that this phenomenon also exists for other hypergraph topologies. 

    \paragraph{Barab\'asi-Albert Hypergraphs}

        It is known that behaviour of the standard DW dynamics is qualitatively similar on the ER, and the  BA networks. \cite{stauffer2004simulation,weisbuch2004bounded}.
        Its  generalization to HOI introduces, nevertheless, some differences. 
        While as for the ER hypergraph, the behaviour of the HIOD in the BA hypergraph differs more and more from the corresponding dyadic model as $k$ increases, this differentiation is stronger than for the ER case:
        already for $k=3$ the polarization plateau completely disappears.
        The size dependence of the order parameter $\avg{S}$ and the variance, still suggest a phase transition, however, the the difference between the values of $\avg{S}$ before and after the transition point is much smaller, than for the ER case.
        
        \begin{figure}[htb]
            \centering
            \includegraphics[scale=1]{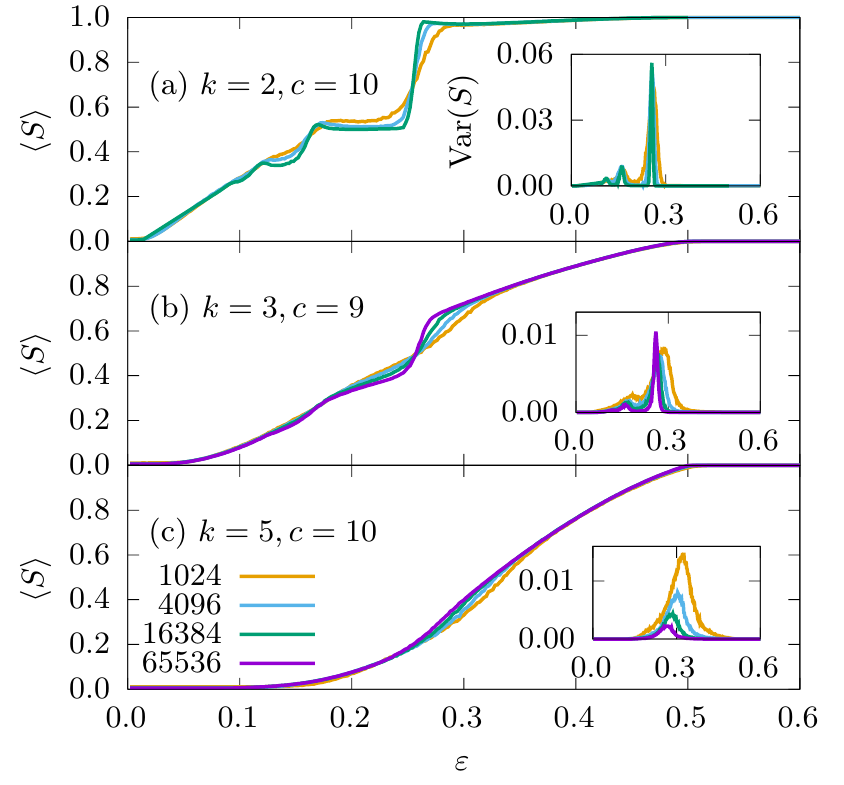}
            \caption{\label{fig:hbc_ba9_s}
                Mean relative size of the largest cluster $\avg{S}$ as a function of the confidence $\varepsilon$ for the HOID model on $k$-uniform BA with $k \in \{2, 3, 5\}$ and an expected mean degree of $c=10$ for different system sizes. The insets show the variance $\mathrm{Var}(S)$, which is sharpening for $k \le 3$ and vanishing for $k = 5$; the same behavior as for the ER.
            }
        \end{figure}
\begin{figure}[htb]
            \centering
            \includegraphics[scale=1]{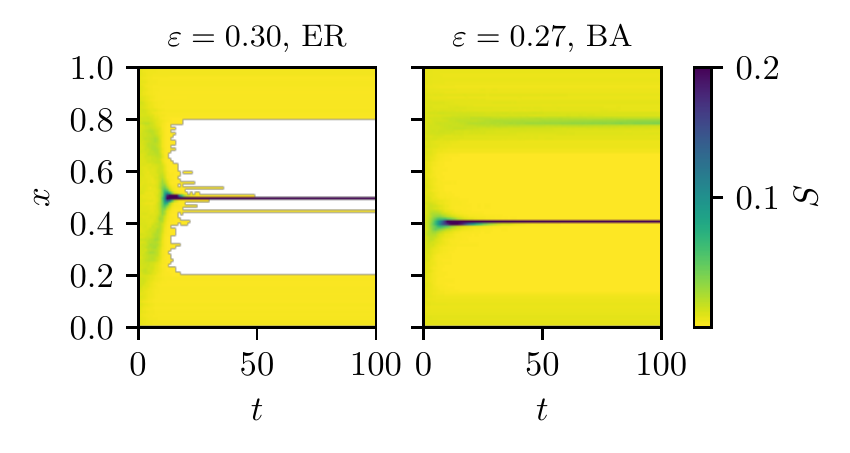}
            \caption{\label{fig:timedev_er_ba}
                Examples of the trajectory of a single realizations of the HOID model on ER ($k=3, c=10$) and BA ($k=3, c=9$) hypergraphs for $N=16384$ close to their transition to consensus. The horizontal axis is time in units of sweeps, i.e., $N$ attempted updates. The dark colors show regions where agents are highly concentrated, light color show regions with only very few agents and white signifies the absence of any agents. The colormap is truncated at $S = 0.2$, to better visualize the small clusters, therefore the darkest shade represents all values $0.2 \le S \le 1.0$.
            }
        \end{figure}

        In order to understand the mechanisms that lead to these different behaviours of the order parameter, we examine the trajectories for both systems  near the corresponding critical confidences $\varepsilon_c$. Fig.~\ref{fig:timedev_er_ba}. 
        shows that while the trajectories for the ER case evolve to consensus by joining two symmetric strands, (left panel of Fig.~\ref{fig:timedev_er_ba}), those of the BA case are asymmetric, with one majoritarian strand that contains a much larger share of agents (right panel of Fig.~\ref{fig:timedev_er_ba}). 
       
       Similar to the ER case, the sharp turns into a smooth crossover for larger values of $k$.

\begin{figure}[htb]
            \centering
            \includegraphics[scale=1]{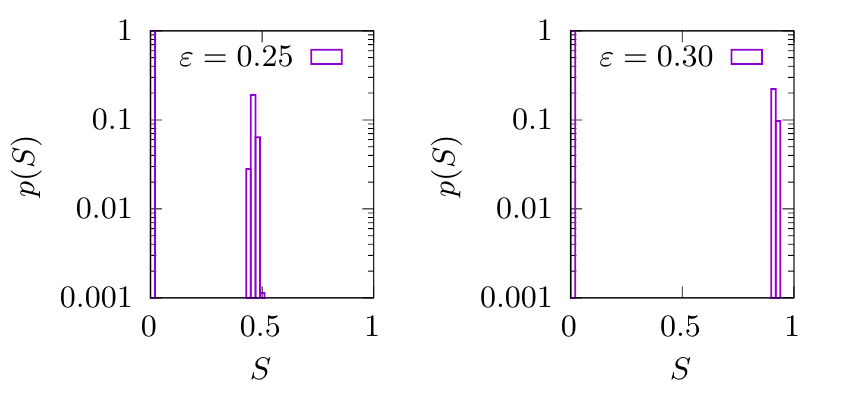}
            
            \includegraphics[scale=1]{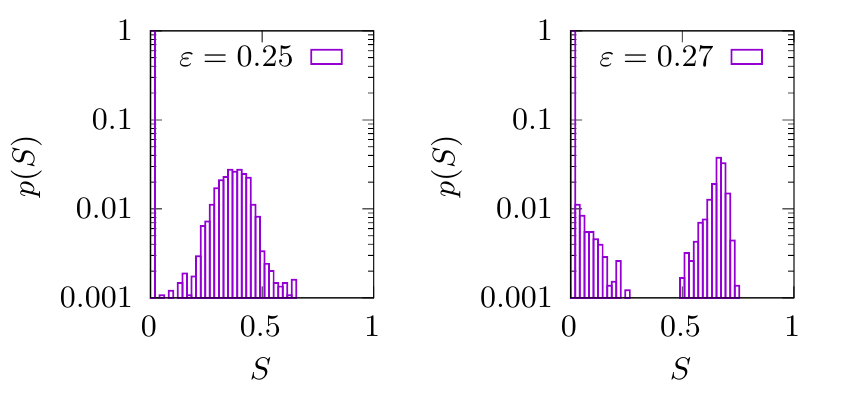}
            \caption{\label{fig:size_distrib_ER_BA}
                Cluster size distribution of systems of size $N=65536$ below and above the corresponding $\varepsilon_c$. Top panel:  ER $3$-uniform hypergraph, $\varepsilon_c= 0.277(1)$. Bottom panel:  BA $3$-uniform hypergraph, $\varepsilon_c = 0.257(1)$.
            }
        \end{figure}

      Figure ~\ref{fig:size_distrib_ER_BA}, illustrates the same phenomenon at the final state, where one can observe  that the    cluster size distributions before and after the transition  are very different for ER and BA hypergraphs. The  almost isolated peak at $S=0.5$ of the ER hypergraph before the transition shows that polarization involves two equally populated strands that join into a single one after the transition ($\varepsilon=0.3$). On the contrary  for  the BA hypergraph, the distribution is broad, around $S \approx 0.4$, before the transition. Moreover, 
      letting aside the very small clusters ($S \leq 0.1) $, it looks quite symmetrical around the peak which  still indicates polarization although with the existence of branches that could be unequally populated. At the transition, this distribution  presents a sharper peak at  $S \approx 0.7$ showing that one of the two strands has gathered more agents than the other.  
A heuristic argument to explain these differences could be related to the very high degree nodes (hubs) that are found, by construction, in the BA hypergraph. As a hub belongs to many hyperedges, it is likely to get unblocked in some of the hyperedges it belongs  to. As it interacts in one of those hyperedges, its opinion evolves, allowing for the unblocking of the other hyperedges it also belongs to, and therefore, attracting all the nodes belonging to those hyperedges to a common opinion. Other nodes, with less connectivity and not directly connected to the hub, are  less likely to grow a large cluster. In the ER on the other hand, there are no hubs, and it is possible to observe several nodes of relatively  high connectivity distributed around the network and not directly connected. They  can grow clusters independently around different opinions before joining into a single strand at higher confidence values. 

    \subsubsection{Non-uniform Hypergraphs}

As shown on the previous sections, the behaviour of the system for a given connectivity strongly depends on the size of the hyperedges. Therefore it is interesting to study what happens when the system contains hyperedges of different sizes. As we have seen that large hyperedges are more prone to remain blocked one could expect that the dynamics is lead by the smaller, non blocked ones. To further investigate this point we study a system combining hyperedges of $k=3$ and $k=5$ which have revealed different behaviours in the uniform hypergraphs. We study  two different ways of combining these hyperedges of different sizes: (a)  same average connectivity for edges of different size  (b)  same average number of hyperedges of both kinds.  
        
        From eq.~\ref{eq:er_c} and, knowing that the average number of hyperedges of size $k$ is 
        $M_k = \binom{N}{k} p_k $, one obtains the ratio of hyperedges of each kind  $M_3 = 5/3 M_5$ for the case (a), where we have fixed, $c_3 = c_5 =5 $ such that $c= c_3 + c_5 = 10$, for comparison with previous results.  For case (b), fixing again $c=10$ one obtains the mean degrees of the hyperedges of both sizes as $c_3=30/8$ and $c_5=50/8$, which are on the order of the sparse hypergraphs previously considered. 

        \begin{figure}[htb]
            \centering
            \includegraphics[scale=1]{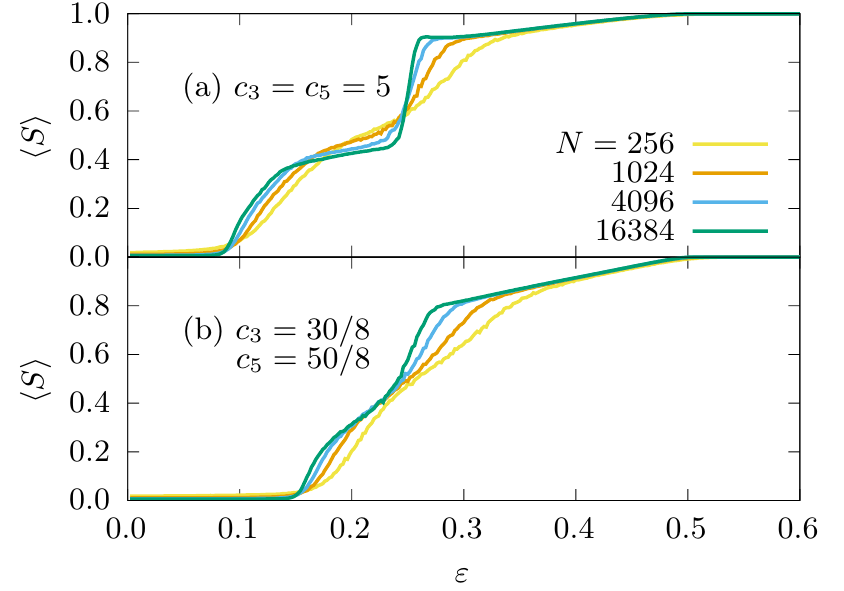}
            \caption{\label{fig:hbc_er10_k3_5_s}
                Mean relative size of the largest cluster $\avg{S}$ as a function of the confidence $\varepsilon$ for the HOID model on heterogeneous hyper-ER graphs with a connectivity of $c = 10 = c_3+c_5$, where the connectivity $c_3$ is caused by $k=3$ hyperedges and $c_5$ by $k=5$ hyperedges. In case (a) both types of hyperedges cause the same connectivity $c_3=c_5=5$. In case (b) there are on average equal numbers of both hyperedges, i.e., $c_3=30/8$ and $c_5=50/8$.
            }
        \end{figure}

        Indeed, we observe that in case (a) the shape of the curve is a slightly smoothed version of the $3$-uniform case without a hint for fundamentally new behavior, showing that the $k=3$ hyperedges dominate the behaviour. This is not surprising as they are more numerous and less susceptible to be blocked. On the other hand, when the system contains the same number of hyperedges of both kinds one does not observe the predominance of the $k=3$ behaviour, instead the curves look similar to those of the 'intermediate' $4$-uniform case. Interestingly, the same qualitative behaviour is observed when mixing dyadic edges with $k=4$ hyperedges, both randomly distributed.
        
        We therefore conjecture that when the average number of hyperedges of different sizes is the same, non-uniform hypergraphs do not behave too differently from the uniform hypergraphs with $k$ in the same range, with a behaviour that `interpolates' between the two considered uniform cases. In particular, we do not observe as could have been expected, the smaller $k$ dominating the behavior in this case, in spite of the property of smaller hyperedges to be much less susceptible to blocking. This is an interesting finding, because it gives a hint of the behaviour of heterogeneous hypergraphs: hyperedges of size $k$ may dominate the behaviour when they are many more than the others, however when the average number of hyperedges is similar for all sizes, the expected behaviour would be similar to a uniform hypergraph with an intermediate value of $k$. 

    \subsection{Regular, Spatial Hypergraphs}
    
\begin{figure}[htb]
            \centering
            \includegraphics[scale=1]{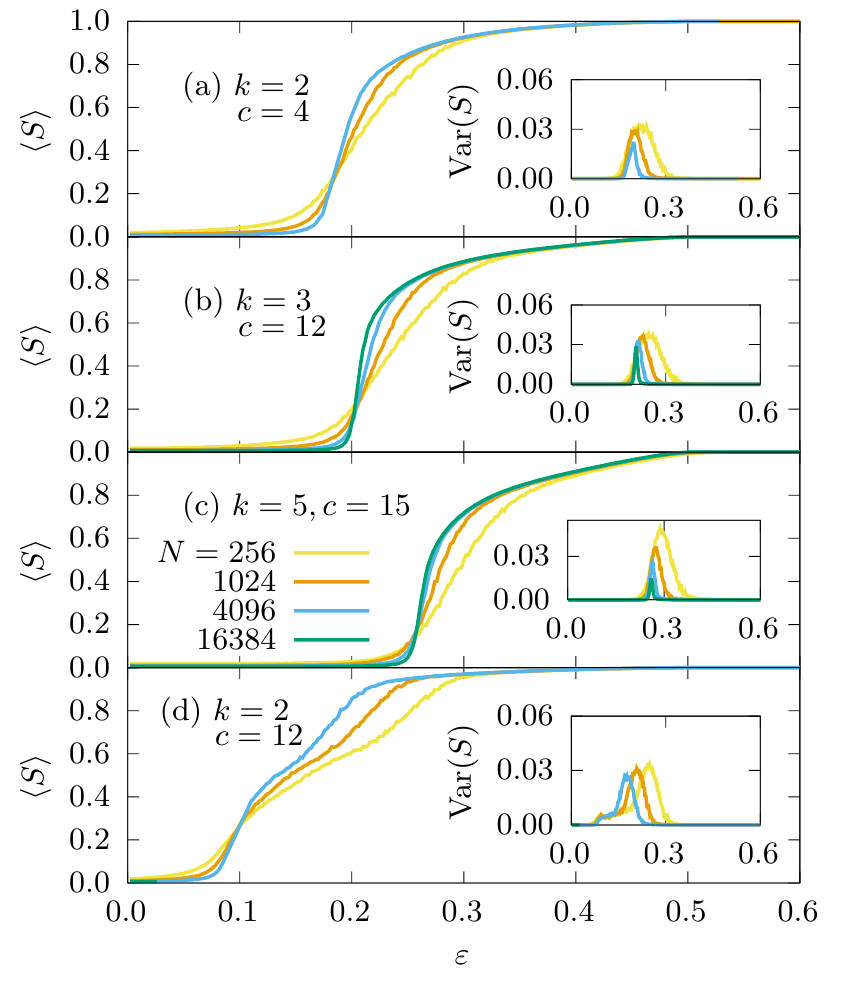}
            \caption{\label{fig:lattice}
                Mean relative size of the largest cluster $\avg{S}$ as a function of the confidence $\varepsilon$ for the HOID model on different lattice-like topologies for different system sizes. Case
                (a) is a nearest neighbor square lattice with $k=2$ and $c=4$, case
                (b) and (c) are the spatial hypergraphs defined in Fig.~\ref{fig:lattice_ex} with $k=3, c=12$ and $k=5, c=15$, and case
                (d) is a square lattice with $k=2$ and up to third nearest neighbor interactions, i.e., $c=12$.
            }
        \end{figure}

        We present here the results for the HOID model on  hypergraphs built in such way that they keep the regularities and spatial symmetries of square lattice, with hyperedges including first, second and third nearest neighbours. 
   In Figure~\ref{fig:lattice} the results on hypergraphs with hyperedges of sizes $k=3$ (panel (b))  and $k=5$ (panel (c)) are compared with the corresponding dyadic DW model in the square lattice with only nearest neighbours (panel (a)) and with third nearest neighbours (panel (c)). Notice that the three cases have  a similar connectivity, $c \approx 12, 15$, which is also of the order of the sparse random hypergraphs studied in section~\ref{subsubsec:uniform}

     Figure.~\ref{fig:lattice} shows that the behaviour of regular hypergraphs is completely different from the random case: no polarization is observed for both $k$ values, and there is a sharp transition from complete fragmentation to consensus, that seems continuous (panels (b) and (c)). Moreover  unlike for random hypergraphs,  no crossover to a smooth size independent behaviour for $k=5$ is found.  
     
      Interestingly, while a longer reach of the interactions in a dyadic lattice  favours consensus, lowering the value of $\varepsilon_c$  (panel (d)), 
        it does not for hypergraphs, because reaching further neighbours implies involving larger group sizes which are easier to get blocked. 

        \begin{figure}[htb]
            \centering
            \includegraphics[scale=1]{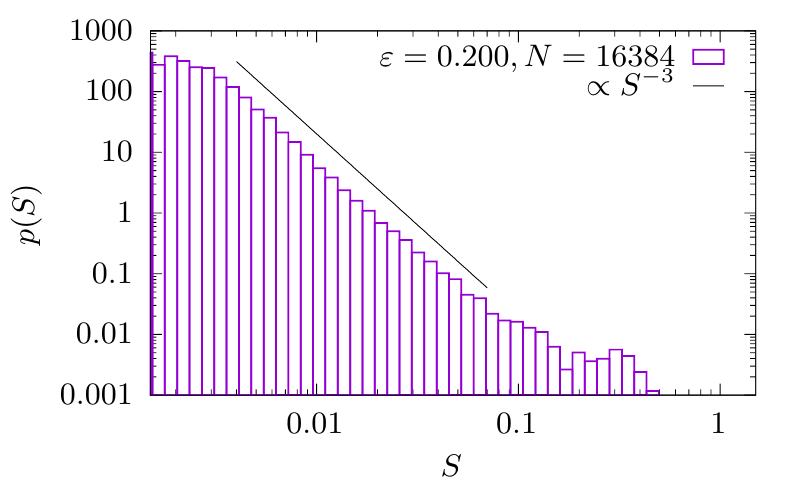}
            \caption{\label{fig:lattice_clustersize}
                Cluster size distribution at $\varepsilon = 0.2$ for the HOID model on a lattice with $k=3$, $c=12$.
            }
        \end{figure}
As  for continuous phase transitions in thermodynamics, we observe a scale free distribution of cluster sizes. This is the indication of a divergent correlation length   in the thermodynamic limit (see  Fig.~\ref{fig:lattice_clustersize}). However we were not able to scale the order parameter, $\avg{S_N}(\varepsilon) = \widetilde{S}\left((\varepsilon - \varepsilon_c) N^{\nu}\right)$  with a unique exponent above and below the transition. 

    \section{Conclusions}
    \label{sec:conclusions}
    
   We generalized the Deffuant model to higher order interactions, where the discussions take place in small groups that cannot be reduced to combinations of pairwise interactions. This generalization involves some hypothesis on how the interaction will take place inside the group. Unlike in Ref.\cite{hickok2021bounded}, where a discordance function facilitating the interaction of large groups is proposed, here we just follow the original DW premises by considering that in order to interact, all members of the group must hold opinions inside the confidence interval. This working hypothesis describes the situation where some agents in the group, by holding an opinion that is very far from the others' may block an otherwise possible compromise. As a consequence, it is more difficult to reach a common opinion as the group becomes larger. At this point it should be noticed that we have generalized the Deffuant model in the case where the interacting agents end up having  the same opinion after the interaction. In other words, the parameter $\mu$ of the original model is set to its maximum value  $\mu=1/2$. This is a  very common choice  (see for ex.~\cite{hickok2021bounded}) for computational reasons, because this  parameter controls the speed of convergence of the model~\cite{deffuant2000mixing}. However care should be taken in the case of  a heterogeneous model, with the agents characterized by different confidences, as it has been shown that, in this case the characteristic  time scales are different for different confidences and modify the outcomes of the dynamics~\cite{schawe2020open}.  
   
   The natural tool to model group interactions are hypergraphs. We therefore perform large scale simulations of this HOID model in sparse hypergraphs with different hyperedge distributions. We observe that the fact of introducing hyperedges completely changes the dynamics with respect to the DW model in the corresponding network topology. 
   One reason for this is the blocking effect of the hyperedges, which requires higher confidences to overcome fragmentation (the domain of $\varepsilon$ values where $\avg{S}(\varepsilon) = 0$ increases with $k$). 
   
   The most interesting result of the inclusion of hyperedges, is that above a certain size of the groups the sharp transition to consensus, well known from the original Deffuant model, changes to a smooth crossover. In other words, when discussions take place in groups, a small decrease in the confidence will not trigger a sharp disruption of the society from consensus to polarization but a slow decrease of the amount of individuals sharing the same opinion.  Since such crossovers are generally preferable to sharp transitions for the stability of real societies, this fundamental mechanism could be of further interest. The counterpart is that, in order to unblock larger hyperedges, larger confidences are required to leave the fragmentation region. 
   When the hypergraph becomes dense the phase transition still holds for  hyperedges of $k=6$, but it is nevertheless less sharp than for smaller hyperedges.  
   
   Furthermore, the introduced model shows a richer behavior on different hypergraph ensembles: while the dyadic Deffuant model behaves qualitatively in the same way on ER or BA networks, its generalization to hypergraphs shows different   agents' opinion trajectories for ER and BA hypergraph. 
   
   It is worthwhile stressing the importance of studying finite size effects for those systems. 
 As system  size increases, new phenomena, absent for the smaller ones, often appear. In other cases the independence of the order parameter with the size allows to distinguish a phase transition from a smooth crossover.  We note that many studies on opinion dynamics do not look at the size dependence at all or study only fairly small sizes, such that similar fundamental differences for other systems might have been overlooked in the past.

      We also show that if the hypergraph is not homogeneous, the expected dominance of small hyperedges (due to the fact that larger ones are more likely to be blocked) does not occur , unless they are significantly majoritarian. If the average number of hyperedges of different sizes is the same, the system behaves similar to a uniform hypergraph case with an intermediate $k$ value. As a consequence, heterogeneous hypergraphs, which are nearer to real societies, will mostly lead to a smooth crossover rather than to a sharp transition. 
   
   Interestingly, spatially structured hypergraphs, unlike random ones,  seem to show the same behavior as dyadic lattices. However,  increasing the reach of the interaction in the hypergraphs shows the opposite behaviour than doing so in lattices. While there are no qualitative changes in hypergraphs, beside the expected increase of $\varepsilon_c$, a higher reach in dyadic lattices promotes consensus. This is a consequence of the competing effects on including further neighbours in the hypergraph, on the one hand it increases the reach but on the other, by increasing the size of the group, it enhances the possibility of blocking hyperedges.  
   
   Finally, interpreting each hyperedge as a social group, leads to the obvious extension to the dynamics by granting the agents the ability to leave groups to join another group. The departure  could be triggered by some measure of the \emph{frustration} of not being able to reach consensus within a group. This way of  unblocking larger edges, which models behaviour in real systems,  is the subject of forthcoming work.\\

 \section*{  \textbf{Data and Code Availability} }

Data on cluster configurations for the Hyper Bounded Confidence model are available at \href{https://doi.org/10.5281/zenodo.5026816}{doi.org/10.5281/zenodo.5026816}.
     
Data on cluster configurations for the Deffuant model are available at 
\href{https://doi.org/10.5281/zenodo.4701047}{doi.org/10.5281/zenodo.4701047}. 

The complete code is available at  https://github.com/surt91/hk \\

 \section*{ \textbf{Authorship} }

Both authors jointly, conceived the research questions,  designed the research protocol, analysed  the results and wrote the manuscript. H.S created and executed the code.\\

 \section* {\textbf{Competing interests}} 

The authors declare no competing interests.

    \section*{Acknowledgments}
        The authors acknowledge the OpLaDyn grant obtained in the 4th round
        of the Trans-Atlantic Platform Digging into Data Challenge (2016-147 ANR OPLADYN TAP-DD2016)
        and Labex MME-DII (Grant No. ANR reference 11-LABEX-0023).

    \bibliography{lit}

\end{document}